\documentclass[prl,a4paper,twocolumn,amsmath,superscriptaddress]{revtex4}
\usepackage{graphicx}

\begin{document}

\title{Non-monotonic density dependence of the diffusion of DNA
fragments in low-salt suspensions}

\author{Mathieu~G.~McPhie}
%\email{m.mcphie@fz-juelich.de}%%
\affiliation{Institut f\"ur Festk\"orperforschung, Forschungszentrum
J\"ulich, D-52425 J\"ulich, Germany}%%

\author{Gerhard~N\"agele}
\email{g.naegele@fz-juelich.de}%%
\affiliation{Institut f\"ur Festk\"orperforschung, Forschungszentrum
J\"ulich, D-52425 J\"ulich, Germany}%%

\date{\today}

\begin{abstract}

The high linear charge density of 20-base-pair oligomers of DNA is
shown to lead to a striking non-monotonic dependence of the
long-time self-diffusion on the concentration of the DNA in low-salt
conditions. This generic non-monotonic behavior results from both the
strong coupling between the electrostatic and solvent-mediated
hydrodynamic interactions, and from the renormalization of these
electrostatic interactions at large separations, and specifically
from the dominance of the far-field hydrodynamic interactions caused by
the strong repulsion between the DNA fragments.

\end{abstract}


\maketitle

%%%%%%%%%%%%%%%%%%%%%%%%%%%%%%%%%%%%%%%%%%%%%%%%%%%%%%%%%%%%%%%%%%%%%
%%%% INTRODUCTION %%%%%%%%%%%%%%%%%%%%%%%%%%%%%%%%%%%%%%%%%%%%%%%%%%%

With the increasing importance of biophysics there is a greater
overlap between the subjects of the physics of biological molecules
and colloidal physics, which deals with small particles in
suspension. Many of the techniques that have been developed in
colloidal physics are directly applicable to biological molecules,
such as proteins, and cells. DNA in particular is a very
interesting biomolecule which exhibits a wide range of behaviors,
due to its interactions with proteins and enzymes, but also due to
its physical characteristics. A particular feature of DNA that is of
interest to colloidal scientists is its high linear charge density
\cite{wilk-jcp,qiu-prl}.
The interaction energy between DNA molecules and the structure of
suspensions of DNA fragments have been shown to be strongly affected
by this charge density. Consequently, other effects associated with
a strong charge, such as electrolyte screening, electrolyte
friction, charge condensation, charge inversion and like-charge
attraction in the presence of multivalent salt ions, have also been
predicted and investigated
\cite{mcphie-dense,pian-barbosa-levin,qiu-prl-2007,Allahyarov:04}.
Several investigations have considered the effect on
the structure due to the renormalization of charge
\cite{qiu-prl,rojas-ochoa-prl}. Little work, however, has been
performed on the interesting consequences of a high charge on the
dynamics of these molecules.

Highly charged colloids
present an entirely different paradigm of particle interactions
to the classical
hard-sphere model. Well-known features of a low-salt suspension of
charged colloids are the low osmotic compressibility of the
suspension, and that the mean distance between the particles scales
with the inverse of the cube root of the colloid concentration over
a wide range of concentrations \cite{banchio-jcp}. This
scaling corresponding to the formation of a correlation hole
around the particles, leads to the domination of
the far-field over the near-field hydrodynamic interactions (HI),
and therefore to an altered dynamic behavior which is unlike that
of hard-sphere colloids.
%This suppression of near-field
%HI should not be confused with the picture of
%HI screening which has recently been comprehensively
%refuted in fluids of charged colloids \cite{gn-prl}.

Wilk \textit{et al.}~\cite{wilk-jcp} have measured
the long-time translational self-diffusion coefficient, $D^L = \left(D^L_{\parallel} + 2\;\! D^L_{\perp} \right)/3$, of
isotropic dispersions of 20 base pair
oligomers of DNA by fluorescence correlation spectroscopy for
various salt concentrations. DNA is a suitable molecule for studying
effects resulting from electrostatic and hydrodynamic coupling due
to its large linear charge density of approximately
$-2e/3.4$~{\AA}. The 20-mer DNA are almost perfectly monodisperse,
rigid cylindrical rods with length $L = 6.8$~nm and diameter $d =
2$~nm (aspect ratio of $3.4$),
%molecular weight $M_w = 13022$ g mol$^{-1}$,
a bare valency of $Z=-42$ in neutral or basic \textit{p}H solutions
and the translational free diffusion coefficient of $D^0 = 1.07 \times
10^{-6}$~cm$^2$~s$^{-1}$. The long-time coefficient, $D^L$, was
found to have an unexpected non-monotonic concentration dependence
in low-salt conditions.

In this letter we describe a versatile theoretical scheme that we have
developed for the calculation of $D^L$ in colloidal systems
\cite{kollmann,mcphie-finitesize,mcphie-dense}. This
scheme includes the long-range far-field part of the HI between the
particles, which dominates in low-salt suspensions. We will show
that in combination with colloid charge renormalization, this scheme
successfully describes the non-monotonic dependence of
$D^L(\phi)$ on the macroion volume fraction $\phi$. This effect is of general
importance and may be observed in any dispersion of colloids or biomolecules
where long-range repulsive interactions are prevalent.
The non-monotonicity in $\phi$ is unusual
since it requires a delicate interplay
of HI and electrosteric repulsions over a sufficiently
broad concentration range. In contrast,
a non-monotonic $\phi$-dependence
is not uncommon for transport properties such as the primary electroviscous
coefficient, $p(\phi)$, associated with the
suspension viscosity $\eta$ \cite{Carrique:05}, and the collective
diffusion coefficient $D^c$ \cite{banchio-jcp}. Moreover,
the electrophoretic mobility $\mu$ \cite{lobaskin-prl}
as well as $1/\eta$ and
$D^L(\phi =0)$ exhibit a minimum as a function of the
electrolyte concentration.
The minimum in $D^L(0)$ is found for globular
 macroions \cite{kollmann,Geigenmueller:84}, and
also for semi-flexible charged polymers as shown recently
in experiments and
simulations \cite{netz-epl}.
In all cases considered, HI play a decisive role, e.g.,
the maximum of $D^c(\phi)$ arises from a balance
of the slowing HI and the speed-up of density relaxations
caused by the electrosteric repulsion. The maximum in $p(\phi)$ at intermediate salinity
arises from a competition between the velocity gradient
inside the macroion double layer that grows with $\phi$,
and the shrinking double layer distortion \cite{Carrique:05}.
Ignoring HI can lead to nonphysical results such as the
failure to predict the maximum in the electrophoretic mobility of
a short polyelectrolyte chain as a function of the
monomer number \cite{Grass:08,Frank:08}.

Due to the increasing power of computers, substantial
progress has been made in the simulation of transport
properties in non-dilute charged colloidal dispersions.
Formerly, simulations that include the electrokinetic effect of electrolyte
ions have been difficult due to the large asymmetry between these
components, both in size and charge. These simulations have focused
largely on the challenging problem of electrophoresis
\cite{lobaskin-prl,tanaka-epl,kim-prl}, frequently
used for particle characterization but for which a complete theory
in dense systems, in which there is strong overlap of the electrical
double layers, is still lacking. Our theoretical scheme is therefore
also presented as a significant step in developing a versatile
statistical mechanical tool to describe the electrokinetic transport
in non-dilute suspensions.

%%%%%%%%%%%%%%%%%%%%%%%%%%%%%%%%%%%%%%%%%%%%%%%%%%%%%%%%%%%%%%%%%%%%%
%%%% MODE COUPLING THEORY AND ELECTROLYTE FRICTION %%%%%%%%%%%%%%%%%%

Our scheme is based on an exact memory equation for the
self-intermediate scattering function of colloids undergoing
overdamped Brownian motion. The irreducible memory function in
this equation is approximated using the idealized mode-coupling
theory for Brownian fluids, with the important extension of
including the HI between all ionic species.
Instead of solving the mode-coupling equations fully
self-consistently, which would be a very challenging numerical task
in the presence of HI, we use a simplified solution scheme which
retains analytical simplicity and yields only small differences in
the numerical results \cite{nagele-baur}.

According to our scheme, the long-time coefficient $D^L$ of a
non-dilute suspension is given by the Stokes-Einstein-like relation
\cite{mcphie-dense}
\begin{equation}\label{eq:CFandEF}
\frac{D^L}{D^0} = \left[ 1 + \frac{\Delta\zeta^{CF}}{\zeta^0} +
\frac{\Delta\zeta^{EF}}{\zeta^0} \right]^{-1},
\end{equation}
which includes, in addition to the colloid-solvent friction $\zeta^0
= 6\pi \eta^0 a$, where $\eta^0$ is the solvent viscosity and $a$ the
colloid radius, a colloid friction (CF) and an electrolyte friction
(EF). The colloid friction arises from the
microion-averaged electrosteric and HI
between the colloids, and is present even when the
microion degrees of freedom are ignored. It is given in our scheme
by \cite{mcphie-dense}
\begin{equation}\label{eq:fric-cf}
\frac{\Delta\zeta^{CF}}{\zeta^0} = \frac{n}{6\pi^2} \int_0^\infty dk
\, k^2 \frac{\left[ h(k) - \frac{1}{D^0} h^d(k) \right]^2} {2 + n
\left[ h(k) + \frac{1}{D^0} h^d(k) \right]},
\end{equation}
where $n$ is the colloid number density, $h(k)$ is the total
correlation function of the colloids, $h^d(k)$ is the distinct
hydrodynamic function of the colloids, and $D^0$ is the
free-diffusion constant. This expression for
$\Delta\zeta^{CF}$ only requires the colloid static
structure factor $S(k) = 1 + n h(k)$.
Eq.~\eqref{eq:fric-cf}
is the zeroth order term
in the expansion of the total long-time friction
coefficient, $\Delta\zeta = \Delta\zeta^{CF} + \Delta\zeta^{EF}$,
in terms of the colloid-microion mobility ratio $D^0/D_i^0$.
Eq.~\eqref{eq:CFandEF} states that the extra friction
due to the fast kinetics of the mobile salt ions, $\Delta\zeta^{EF}$,
is given by the difference of the total friction
coefficient and $\Delta\zeta^{CF}$
\cite{mcphie-dense}. The EF is due to
the non-instantaneous relaxation of the microionic
atmosphere.
We have derived a
simple expression for the long-time EF
contribution, valid for the case when the free diffusion
coefficients of the various salt ion species, $D^0_i$, are much
greater than the colloid free diffusion coefficient. This expression
is
\begin{align*}
&\frac{\Delta\zeta^{EF}}{\zeta^0} = \frac{2}{3 \pi^2} \sum_{i=1}^m
n_i \frac{D^0}{D^0 + D^0_i} \int_0^\infty dk \, k^2 \times \\
&\left\{ \frac{ \left[ 1 + nh(k) \right] \frac{1}{D^0} h^d_{ci}(k) -
h_{c i}(k) \left[ 1 + n\frac{1}{D^0} h^d(k) \right] }{ 2 + n \left[
h(k) + \frac{1}{D^0} h^d(k) \right] } \right\}^2,
\end{align*}
where the sum goes over all microion species of number density
$n_i$, and where $h_{c
i}(k)$ and $h^d_{ci}(k)$ are the partial total correlation and
partial distinct hydrodynamic functions between the microions and
the colloids.
The EF contribution to $D^L$ is
significant in very
dilute systems but is negligible when the mobility
difference between the colloidal spheres and the microions is large,
and when $\phi$ is increased.
The second finding is attributed to the enhanced homogenization of the
electrolyte background with increasing $\phi$ \cite{mcphie-dense}.

%%%%%%%%%%%%%%%%%%%%%%%%%%%%%%%%%%%%%%%%%%%%%%%%%%%%%%%%%%%%%%%%%%%%%
%%%% EFFECTIVE POTENTIAL AND CHARGE %%%%%%%%%%%%%%%%%%%%%%%%%%%%%%%%%

Since the dynamic effect of the microions is small, we can
simplify the problem by considering the colloids as
interacting with an effective pair potential. In the resulting
one-component model of weakly charged colloids, the effective
interaction between colloids of radius $a$ and bare valency
$Z_\text{bare}$, follows the repulsive part of the DLVO
potential \cite{DLVO},
\begin{align*}
\beta u(r) = L_B Z_\text{bare}^2 \left( \frac{e^{\kappa a}}{1 +
\kappa a} \right)^2 \frac{e^{-\kappa r}}{r}, && r > 2a
\end{align*}
where $L_B = e^2/(4\pi\epsilon k_B T)$ is the Bjerrum length in a
solvent of dielectric constant $\epsilon$ and $\kappa$ is the
inverse Debye screening length determined by the concentration of
added salt ions and monovalent counterions. In a
1:1 electrolyte solution, $\kappa^2 = 4\pi L_B \left( 2 n_s + n
|Z_\text{bare}| \right)$. For strongly charged colloids where $L_B
|Z_\text{bare}|/a > 1$, this potential is
still suitable but only with the charge and
screening parameter replaced by an effective charge, $Z_\text{eff}$,
and screening parameter, $\kappa_\text{eff}$, due to the
condensation of counterions near the colloid surfaces. There exist
several schemes for the
calculation of these effective quantities, and these have recently
been of considerable interest. Those mostly used
are the cell model approximation of Alexander \textit{et
al.}~\cite{alexander}, and the renormalized
jellium approximation (RJA) \cite{pian-trizac-levin}.

The RJA for the effective macroion charge in a closed
suspension with a fixed salt concentration, as opposed to a system
in contact with a reservoir, involves numerically solving the
Poisson equation for the mean electric potential $\psi(r)$ of a
single colloidal sphere, surrounded by a Boltzmann-distributed
microion cloud and a uniform negatively-charged background of charge
density, $n Z_\text{eff}$, that describes the jellium representing
the other macroions. The resulting equation for the reduced
potential, $y(r) = \beta e \psi(r)$, is
\begin{equation} \label{eq:jellium}
\nabla^2 y(r) = 4\pi L_B \left[ 2 n_s \sinh(y(r)) + n Z_\text{eff}
(e^{y(r)} - 1) \right]
\end{equation}
with the boundary conditions $y(\infty) = 0$, $y'(\infty) = 0$, and
$y'(a) = - L_B Z_\text{bare}/a^2$. The numerical
solution of Eq.~\eqref{eq:jellium} is
asymptotically matched to the solution of the linearized equation,
$\nabla^2 y_\text{lin}(r) = \kappa_\text{eff}^2 y_\text{lin}(r)$,
where $\kappa_\text{eff}$ is the effective screening parameter
$\kappa_\text{eff}^2 = 4\pi L_B \left( 2 n_s + n |Z_\text{eff}|
\right)$. The effective charge comes from the linearized solution at
the inner boundary, $y'_\text{lin}(a) = - L_B
Z_\text{eff}/a^2$. Since the effective charge also appears in
Eq.~(\ref{eq:jellium}) this solution procedure must be iterated
until self-consistency in $Z_\text{eff}$ has been established
\cite{pian-trizac-levin}.

With the so-determined interaction potential, $S(k)$
is calculated using standard integral
equation theories. For this study, we use the
rescaled mean spherical approximation (RMSA) \cite{rmsa},
known to be a reasonably accurate theory of the structure of charged
colloids.

%%%%%%%%%%%%%%%%%%%%%%%%%%%%%%%%%%%%%%%%%%%%%%%%%%%%%%%%%%%%%%%%%%%%%
%%%% DNA PREPARATION, ETC. %%%%%%%%%%%%%%%%%%%%%%%%%%%%%%%%%%%%%%%%%%

Eq.~(\ref{eq:fric-cf}) has been developed for spherically symmetric
particles and is therefore not directly applicable to rodlike
particles in high-salt suspensions. On the other hand, in the
low-salt suspensions, as measured by Wilk \text{et al.}, the
microstructure is mostly determined by the long-range electrostatic
monopole term which is spherically symmetric.
Therefore, we treat the DNA fragments as spheres with an effective
radius, $a_\text{eff}$. This radius appears in the solution of the
effective charge and thus in the calculated structure factor.
The largest effect of $a_\text{eff}$ however, is to determine
the scaling used to map the volume fraction dependent calculations
onto the weight concentration dependent measurements by
$c(\textrm{g/L}) = 3 \phi M_w/(4\pi a_\textrm{eff}^3 10^3 N_A)$,
where $M_w = 13022$ g mol$^{-1}$ is the molecular weight of the DNA
fragments.
The effective radius resulting from the Stokes-Einstein relation
applied to the measured diffusion coefficient at
infinite dilution is $a_\text{eff} = 2.0$~nm. However, since the excluded volume
interaction of the effective spheres is influential at
high salinity only, from comparing our results with the high-salt measurements of
Wilk \textit{et
al.}, we have determined $a_\text{eff} = 3.4$~nm $ = L/2$ to provide the
best overall fit, independent of the salt concentration.
An effective radius of half the molecular
length has some parallels with the excluded volume calculations of
rod-like particles in the isotropic state.
Since the bare charge is given by the number of ionizable groups on the DNA molecule, $a_\text{eff} = L/2$
is the only adjusted parameter.

%%%%%%%%%%%%%%%%%%%%%%%%%%%%%%%%%%%%%%%%%%%%%%%%%%%%%%%%%%%%%%%%%%%%%
%%%% EFFECTIVE CHARGE %%%%%%%%%%%%%%%%%%%%%%%%%%%%%%%%%%%%%%%%%%%%%%%

\begin{figure}[tb]
  \centering
  \includegraphics[scale=0.7]{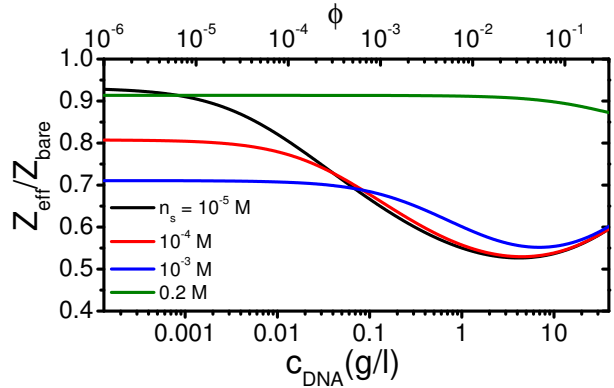}
  \caption{Effective charge of 20-mer DNA
  ($Z_\textrm{bare} = -42$), determined via the RJA, as a function
  of DNA and salt concentrations.
  \label{fig:eff-charge}}
\end{figure}

The effective charge for our model of the  DNA fragments,
calculated via the RJA in a closed system as in the experiment, is
presented in Fig.~\ref{fig:eff-charge}.
For low-salt systems, $Z_\text{eff}$ shows a non-monotonic
dependence on $\phi$.
At very low $\phi$, there is also a non-monotonic dependence of
$Z_\text{eff}$ on the salt concentration (c.f., Fig. 1), with the limit that
$Z_\text{eff} \to Z_\text{bare}$ when $n \to 0$ and $n_s \to 0$.
This zero-$\phi$ non-monotonicity of $Z_\textrm{eff}$ is also
seen in its expansion in terms of the bare
charge $Z_\text{bare}$.
%
%\begin{align*}
%&\frac{Z_\text{eff}}{Z_\text{bare}} = 1 - \frac{(L_B Z_\text{bare}
%\kappa)^2}{6 (1 + \kappa a)^3} \big\{ 1 + 2 (1 - \kappa a)
%e^{4\kappa a} E_1(4\kappa a) \\ &- (1 + \kappa a) e^{2\kappa a}
%E_1(2\kappa a) \big\} + O(Z_\text{bare}^4).
%\end{align*}
%
%where $E_1(x)$ is the exponential integral.
%
%The right-hand-side of this expression has a minimum at $\kappa a
%\approx 1$.
%
If $Z_\text{bare}$ is sufficiently large, $Z_\text{eff}$
becomes independent of the bare charge. This is the so-called
saturated effective charge, $Z_\text{eff}^\text{sat}$. In our system,
$|Z_\text{bare} L_B/a_\text{eff}| \approx 9$ which is well
into the non-linear regime, but $Z_\text{eff}$ is less
than 66\% of $Z_\text{eff}^\text{sat}$, so
that saturation is not yet reached.
Even though the $\kappa_\text{eff}$ depends on $Z_\text{eff}$, it
shows no non-monotonic $\phi$-dependence.

%%%%%%%%%%%%%%%%%%%%%%%%%%%%%%%%%%%%%%%%%%%%%%%%%%%%%%%%%%%%%%%%%%%%%
%%%% LONG-TIME DIFFUSION %%%%%%%%%%%%%%%%%%%%%%%%%%%%%%%%%%%%%%%%%%%%

\begin{figure}[tb]
\centering
\includegraphics[scale=0.7]{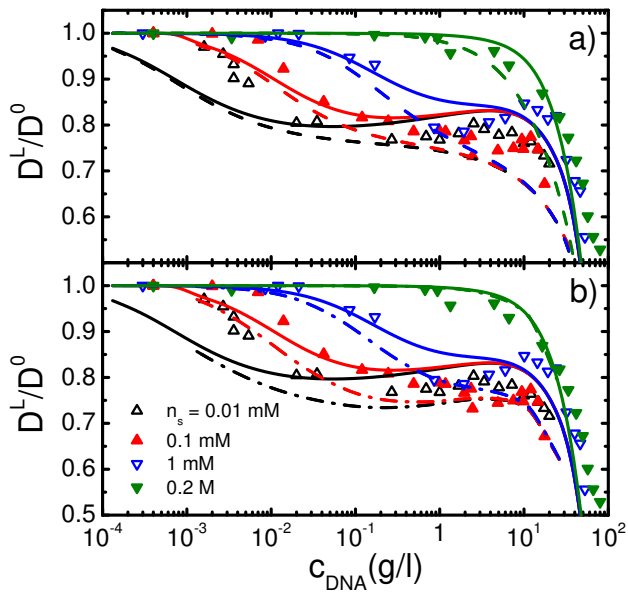}
\caption{Self-diffusion coefficient $D^L$ of the
DNA model vs weight concentration for salt concentrations as
indicated. Symbols are experimental results taken from
\cite{wilk-jcp}.
a) Comparison with theoretical results using $Z_\text{eff}$. Solid
lines are results with HI, dashed lines are without HI.
b) Comparison with theoretical results with HI. Solid lines are
again results using $Z_\text{eff}$, dashed lines are
results with fixed charge $Z=-42$. \label{fig:dna-fcs}}
\end{figure}

The measured $D^L$ of the 20-mer DNA and the comparison
with the results of our spherical model are shown in
Fig.~\ref{fig:dna-fcs}.
In Fig.~\ref{fig:dna-fcs}a, we include the results of our scheme
when far-field HI are included (solid lines) or ignored (dashed
lines), with the values $Z_\text{eff}$ and
$\kappa_\text{eff}$ used in both cases. The non-monotonic
dependence of $D^L$ on $\phi$ in the experiments is also seen in the theoretical
predictions when HI between the particles are included. If the HI
are ignored, $D^L$ shows a monotonic $\phi$-dependence,
for all considered salt concentrations.
In Fig.~\ref{fig:dna-fcs}b, a comparison is made between the
theoretical results for $D^L$, including the effects of HI, where
instead of a $\phi$-dependent $Z_\text{eff}$ (solid lines again)
we use a fixed charge of $Z=-42$ (dash-dotted lines).
For both sets of results a non-monotonic $\phi$-dependence
is seen. The non-monotonicity, however, is much stronger
when the non-constant $Z_\text{eff}$ is used.
This suggests that the non-monotonic $\phi$-dependence of the
experimental $D^L$ results from a simultaneous interplay between the
hydrodynamic enhancement caused by far-field HI, and the
non-monotonic $Z_\text{eff}$ in low-salt systems.
The strong decline in $D^L$ at large $\phi$ seen in
Fig.~\ref{fig:dna-fcs} is due to electrosteric caging
which becomes stronger with increasing $\phi$.
The single-macroion EF effect described in Booth's
theory \cite{Geigenmueller:84} can not explain the
non-monotonicity of $D^L$ since it is significant
only for $\phi < 10^{-4}$ and $n_s \approx 0.01$ M \cite{mcphie-dense},
which is a salt
concentration much larger than those where the DNA-$D^L$
behaves non-monotonically.

The hydrodynamic enhancement of the diffusion of particles repelling
each other over long distances results from the fact that the
dominant far-field HI advect neighboring particles that may
otherwise have hindered the motion of the considered one. Near-field
HI, on the other hand, have the opposite effect of slowing the
diffusive motion.
% They dominate in hard-sphere-like suspensions even
% at low concentration, since particles are likely to be close
%together.
Hydrodynamic enhancement of $D^L$ has been seen before in
suspensions of charged and
magnetically interacting colloidal particles
\cite{zahn-prl,nagele-baur},
and in simulations with HI of charged nano-sized polyions \cite{Jardat:04},
but without a visible non-monotonic $\phi$-dependence. A
non-monotonic $D^L(\phi)$ was found in simulations of
salt-free polyelectrolyte solutions \cite{Chang:02}.
In these simulations, however, HI have been neglected and the
values for $D^L/D^0$ are smaller than $0.1$, which is the
value where the freezing transition of charged spheres
and rods occurs \cite{Simon:93}.

%Another interesting consequence of the present results is that the
%idea of hydrodynamic rescaling
%\cite{medina-noyola-prl,brady,banchio-prl}, successfully used
%in hard-sphere-like suspensions,
%clearly cannot
%work for particles that repel each other over large
%distances. The basic idea behind hydrodynamic rescaling is that the
%time scales over which the direct and hydrodynamic interactions
%affect the motion of the particles are vastly separated, with the HI
%operating only over the short-time dynamics. However, in
%charge-stabilized systems at low salinity the hydrodynamically
%mediated density relaxations are intrinsically coupled with the
%direct interactions.
%
In summary, we have shown that the non-monotonic
concentration dependence of $D^L$ in low-salt suspensions of DNA
fragments can be understood by the influence of far-field HI
and charge renormalization. According to our scheme, the
non-monotonicity of $D^L$ is a generic effect for any low-salt
suspension of strongly charged small colloids or bio-molecules. We
have obtained this result using a simplified mode-coupling scheme.
This scheme is a marked improvement on previous methods, since its
many-component version includes the far-field HI between all ionic
species. Its analytic simplicity allows the study
of electrokinetic phenomena such as the EF effect on
self- and collective diffusion for non-zero concentrations.

\begin{acknowledgments}
This work and the printing of the article was under appropriation of funds from
the Deutsche Forschungsgemeinschaft (SFB-TR6, TP B2).
We acknowledge
A.~Banchio, J.~Gapinski and A.~Patkowski for discussions.
\end{acknowledgments}

%\bibliography{HydroEnhancement}

\end{document}